\documentclass[12pt]{iopart}

\newcommand{\tbar}{\overline{T}}

\begin{document}

%
\title[Comment on cone vacuum energy]{Comment on
 \vspace{\bigskipamount} \\
{\bf  `Wedges, cones, cosmic strings and their vacuum energy'} 
}
\author{S A Fulling and F D Mera} 
\address{Department of Mathematics,
  Texas A\&M University, College Station, TX, 77843-3368 USA} 
\ead{fulling@math.tamu.edu, mera.f@husky.neu.edu}

\begin{abstract}
A recent paper (2012 \emph{J. Phys.\ A} \textbf{45} 374018) is 
extended by investigating the behavior of the regularized quantum 
scalar stress tensor near the axes of cones and their covering 
manifold, the Dowker space.  A cone is parametrized by its 
angle $\theta_1$, where $\theta_1=2\pi$ for flat space. We find 
that the tensor 
components have singularities of the type $r^\gamma$, but the 
generic leading $\gamma$ equals ${4\pi \over \theta_1} - 2$, 
which is negative if and only if $\theta_1>2\pi$, and is a 
positive integer if $\theta_1={2\pi\over N}$.
Thus the functions are analytic in those cases that can be solved 
by the method of images starting from flat space, and they are 
not divergent in the cases that interpolate between those.
As a wedge of angle $\alpha$ can be solved by images 
starting from a cone of angle $2\alpha$, a divergent 
stress can arise in a wedge with $\pi <\alpha \le 2\pi$ but not 
in a smaller one.
\end{abstract}


\maketitle

A recent study \cite{wccsrve} of the Green functions and 
vacuum stress 
tensors of a quantized scalar field in cones and wedges did not 
address the singular behavior (even after point-splitting 
regularization) of those quantities near the axis.
That issue has now become quite pertinent, as a violation of the 
expected relation between torque and energy 
in the wedge \cite{torqanom} or the cone \cite{D13} has been 
discovered and has been tentatively attributed somehow to the 
axial singularity \cite{D13,t1}.
Here we report some further analysis, using \emph{Mathematica}, 
of the regularized energy densities and pressures derived in 
\cite{wccsrve}
for  cones and for the universal cover of cones and wedges,
the Dowker space \cite{D77,D78,D87}.

We review that a cone space-time (also known as the exterior of 
a  cosmic string of infinitesimal diameter) is characterized by 
its angle $\theta_1\,$, which equals $2\pi$ when the cone 
degenerates to Minkowski space.
(The deficit angle, $2\pi-\theta_1\,$, is often used instead.)
Any value $0<\theta_1 \le \infty$ is geometrically possible, 
although in the physical theory of cosmic strings one usually 
assumes $\theta_1 \ll 2\pi$.
The case $\theta_1=\infty$ is the Dowker space.  Green functions 
for any cone can be obtained from the corresponding Green
functions of Dowker space as  infinite periodicity sums.
In turn, a wedge of angle $\alpha$ can be obtained by one 
additional application of the method of images from a cone of 
angle $\theta_1=2\alpha$.  Of course, a physical wedge sitting in 
Minkowski space must have $\alpha\le2\pi$.

The starting point of the calculation is, as usual, $\tbar$, the 
Euclidean Green 
function, or cylinder kernel, of the scalar field in the cone
\cite{L,D78,L87,smith,wccsrve}.
$\tbar$ is \emph{a priori} a  function of two space-time points, 
hence of eight coordinates, but those are reduced to six by 
translation invariance in the temporal and axial directions.
At the end of the calculation the two points are brought 
together, except for one coordinate, say~$z$, maintained as a 
regularization parameter; thus the stress tensor components have 
the form $T_{\mu\nu}(r,\theta,z;\alpha)$ for a wedge of angle 
$\alpha$ and 
$T_{\mu\nu}(r,z;\theta_1)$ for a cone of angle~$\theta_1$.
(To simplify notation, in this paper it is to be understood that 
the ``zero-point'' term, $T_{\mu\nu}(z;2\pi)$, has already been 
subtracted from such objects.)
Formulas for $\tbar$ are summarized in \cite[Sec.~4]{wccsrve};
they are too cumbersome to reproduce here or to manipulate 
further without computer assistance.  Therefore, 
\emph{Mathematica} is used to evaluate the  formulas for the 
expectation values of the components of the stress tensor, whose 
abstract forms in cylindrical coordinates appear in \cite[Sec.~2 
and corrigendum]{wccsrve}.
The most salient feature of the formulas in Sec.~2 is the
appearance of terms 
such as
\begin{equation}
\partial_r{}\!^2 \tbar, \quad 
\bigl.\partial_r \partial_{r'} \tbar\bigr|_{r'=r}\,, \quad
\frac1r\,\partial_r\tbar, \quad
\frac1{r^2}\partial_\theta\!^2
\label{radterms}\end{equation}
requiring two differentiations or divisions 
by a radial coordinate.

Before regularization, $\tbar$ displays two types of 
singularities, those where $r'=r$ and those where either $r$ or 
$r'$ equals~$0$.  The first type is regularized by keeping the 
point separation, $t$ or~$z$, different from $0$.  (In fact, in a 
cone this divergence is then completely removed by the 
subtraction of the flat-space term, but in a wedge or more 
general region there will still be divergences of this type at 
the reflecting boundaries.)  
In \cite{wccsrve} the $t$-splitting was used, but in the present 
work we have used $z$-splitting, because it has become clear 
\cite{EFM,t1} that it is more likely to give physically 
satisfactory results when the cutoff is left finite.
(This change makes no difference in the qualitative 
results we have to report here.)

 Singularities of 
the second type are 
not softened by the point splitting, so it is important to study 
and understand them.  Examination of the formula for $\tbar$ 
reveals that its small-radius asymptotic dependence on $r$ and 
$r'$ collectively involves nonintegral powers, of which 
the most singular is $r^{4\pi / \theta_1}$.
Because of the presence of the terms \eref{radterms}, 
stress tensor components can contain terms with behavior 
as bad as $r^{4\pi / \theta_1 - 2}$.  
This exponent is negative if $\theta_1 > 2\pi$.  
If $\theta_1 < 2\pi$ but not equal to $4\pi /N$, the functions 
are generally  singular (nonanalytic) but nondivergent.  
These expectations are borne out by our detailed 
\emph{Mathematica} investigation of the stress tensor.
The \emph{Mathematica} command \texttt{Series[]} does most of the 
work, but is not able to handle all of the 
asymptotics of our functions, which are both nonanalytic and 
algebraically compound, without some coaxing.
That is, results of some hand calculations needed to be inserted 
at a certain intermediate step; dwelling on the details here 
would be inappropriate.

For the Dowker manifold, the leading singularity in the stress 
tensor is (more easily) seen to be 
$(r \ln r)^{-2}$, which is actually slightly softer 
than the naive  $\theta_1\to\infty$ limit, $r^{-2}$.
It is interesting that  periodic image sums of the Dowker 
Green function are less singular than the function itself.

The maximally singular behavior does not appear in all tensor 
components.
 In the radial and tangential pressures ($T_{rr}\,$, 
$T_{\bot\bot}$ in the notation of \cite{wccsrve}), the 
leading term $r^{4\pi / \theta_1 - 2}$ does appear, with 
coefficient proportional to $1 + 4\beta$, 
where $\beta=\xi-\frac14$ is the curvature coupling parameter of 
the 
scalar field theory.
  It therefore does not 
appear for minimal coupling, $\beta = -\frac14$.  
On the other hand, 
in the energy density and the axial pressure
($T_{00}\,$,~$T_{zz}$) the leading term 
appears with coefficient proportional to $\beta$.  Therefore, it 
does not appear for the calculationally simplest coupling, $\beta 
= 0$  ($\xi = \frac14$).

Finally, observe that $r^{4\pi / \theta_1 - 2}$ is analytic if 
$\theta_1=4\pi/N$ for an integer~$N$.  In particular, this is so 
when $\theta_1=2\pi/N$, the cases where $\tbar$ can be formed 
from the $\tbar$ of flat space by images. This observation 
resolves an apparent paradox, which largely motivated this work.
Because $\tbar(2\pi)$ is nonsingular, $\tbar(2\pi/N)$ must be 
also, and we now see (at least through the lowest orders) that it 
is.  Moreover, numerical 
calculations show no qualitative difference between the vacuum 
stresses of cones with $\theta_1=2\pi/N$ and those with nearby 
real values of $\theta_1\,$, and we now see that those 
interpolating irrational angles display no divergences at the 
axis, because $\theta_1 < 2\pi$.

We have not looked here at the stress tensors of wedges (which 
require manipulating cone Green functions with 
$\theta'\ne\theta$).  However, we anticipate results consistent 
with those for cones. In particular, wedges with angles
$\alpha = \pi /N$ will display no singular behavior at all at the 
axis,
and other angles $\alpha\le \pi$ will display no divergences
there.    
On the other hand, divergent behavior in the 
($z$-cutoff) stress tensor can occur in a wedge with 
$\alpha > \pi$.
Cases like $\alpha = 2\pi / 3$, where the exponent is a positive 
odd integer, should also be nonsingular; we do not know the 
physical or geometrical significance of this fact.

\ack This research is supported by NSF grant PHY-0968269. 
We thank Cynthia Trendafilova for legacy \emph{Mathematica} code.


\Bibliography{00} \frenchspacing


\bibitem{wccsrve} Fulling S A,  Trendafilova C S, Truong  P N 
and  Wagner J 2012
Wedges, cones, cosmic strings and their vacuum energy
\emph{J. Phys.\ A} \textbf{45} 374018;
corrigendum \emph{ibid} submitted

\bibitem{torqanom}  Fulling S A, Mera F D and 
Trendafilova C S 2013
Torque anomaly in quantum field theory
\emph{Phys.\ Rev.\ D} {\bf 87} 047702 



\bibitem{D13} Dowker J S 2013
A note on the torque anomaly
  \texttt{arXiv:1302.1445}


\bibitem{t1}  
    Milton K A,  Kheirandish F,  Parashar P, 
Abalo E K,  Fulling S A,  Bouas J D, 
 Carter H and  Kirsten K 2013
    Investigations of the torque anomaly in an annular sector. I. 
Global calculations, scalar case
\emph{Phys Rev. D} in press [\texttt{arXiv:1306.0866}]

 \bibitem{D77}  Dowker J S 1977
 Quantum field theory on a cone
 \emph{J. Phys. A} {\bf10}  115--124


\bibitem{D78}  Dowker J S 1978
 Thermal properties of Green's 
functions in Rindler, de Sitter, and Schwarzschild spaces
 \emph{Phys. Rev. D} {\bf18} 1856--1860


    \bibitem{D87}  Dowker J S 1987
 Casimir effect around a cone
\emph{Phys. Rev. D} {\bf36}   3095--3101


 \bibitem{L}  Lukosz W 1973
 Electromagnetic zero-point energy shift induced by conducting 
surfaces.~II
 \emph{Z.~Physik} {\bf262}  327--348



\bibitem{L87}  Linet B 1987
 Quantum field theory in the space-time of a cosmic string
 \emph{Phys. Rev. D} {\bf35}  536--539




  \bibitem{smith}  Smith A G 1990
Gravitational effects of an infinite straight cosmic string on 
classical and quantum fields:  Self-forces and vacuum 
fluctuations
 \emph{The Formation and Evolution of Cosmic Strings}, ed
G Gibbons, S Hawking and T Vachaspati (Cambridge:Cambridge)
   pp 263--292

\bibitem{EFM}  Estrada R, Fulling S A and Mera F D 2012
 Surface vacuum energy in cutoff models: Pressure anomaly and 
distributional gravitational limit
  \emph{J.\ Phys.\ A} {\bf 45} 455402 (2012)
 


\endbib
 \end{document}